\documentclass[conference]{IEEEtran}
\IEEEoverridecommandlockouts
\usepackage{amssymb}
\usepackage{amsfonts}
\usepackage{epstopdf}
\usepackage{graphicx}
\usepackage{stmaryrd}
\usepackage{amssymb,amsmath}
\usepackage{multirow,url}
\usepackage{color,cite}
\usepackage{verbatim} 
\usepackage{subfigure}

\begin{document}

\title{\LARGE Timely Status Update: Should ARQ be Used in Two-Hop Networks?
\thanks{The work of J. Feng and H. Pan was supported in part by the Natural Science Foundation of Guangdong Province under Grant 2021A1515012601, in part by the National Natural Science Foundation of China under Grant 62001298, and in part by the Guangdong ``Pearl River Talent Recruitment Program'' under Grant 2019ZT08X603. The work of T.-T. Chan was supported in part by the Faculty Development Scheme (UGC/FDS14/E02/21) and the Research Matching Grant Scheme from the Research Grants Council of Hong Kong, and in part by the Deep Learning and Cognitive Computing Centre, The Hang Seng University of Hong Kong. \emph{(Corresponding author: Haoyuan Pan.)}}
}

\author{
    \IEEEauthorblockN{Jian Feng\IEEEauthorrefmark{1}, Haoyuan Pan\IEEEauthorrefmark{1}, Tse-Tin Chan\IEEEauthorrefmark{2}, Jiaxin Liang\IEEEauthorrefmark{3}}
    \IEEEauthorblockA{\IEEEauthorrefmark{1} College of Computer Science and Software Engineering, Shenzhen University, Shenzhen, China}
    \IEEEauthorblockA{\IEEEauthorrefmark{2} Department of Computing, The Hang Seng University of Hong Kong, Hong Kong SAR, China}
    \IEEEauthorblockA{\IEEEauthorrefmark{3} Department of Information Engineering, The Chinese University of Hong Kong, Hong Kong SAR, China}  
    \IEEEauthorblockA{E-mails: fengjian2020@email.szu.edu.cn, hypan@szu.edu.cn, ttchan@hsu.edu.hk, jiaxin@ie.cuhk.edu.hk}
}

\maketitle

\begin{abstract}
This paper investigates the information freshness of two-hop networks. Age of information (AoI) is used as the metric to characterize the information freshness, defined as the time elapsed since the latest received status update was generated. In error-prone wireless networks, prior studies indicated that Automatic Repeat-reQuest (ARQ) does not help improve the average AoI performance of single-hop networks, because sending a new packet always carries the most up-to-date information (i.e., discarding the old packet). We believe that this observation does not apply to two-hop networks. For example, when a packet transmission fails in the second hop, although a new packet has more recent information, it may require more time to be delivered (i.e., the communication has to restart from the first hop), thus leading to a high AoI. This paper analyzes the theoretical average AoI of two-hop networks with and without ARQ. Specifically, we model the two schemes using Markov chains, from which we derive the average AoI. Our theoretical and simulation results confirm that, unlike single-hop networks, ARQ should be used in two-hop networks to achieve lower average AoI. In particular, when ARQ is used, the successful decoding probability of the second hop has a greater impact on the average AoI than that of the first hop. Overall, our findings provide insight into the ARQ design for two-hop timely status update systems.

\end{abstract}

\section{Introduction}
\label{1}
The Internet of Things (IoT) technologies have substantially promoted the development of machine-type communications (MTC) in 5G communication networks. In many real-time MTC scenarios, such as autonomous driving, telemedicine, and intelligent transportation, timely status updating is of paramount importance. For example, in automatic driving, the status collected by multiple sensors (e.g., real-time locators and lidars) needs to be quickly delivered and integrated for decision-making and control; otherwise, out-of-date status could lead to traffic accidents.

Age of information (AoI) was first proposed in \cite{Kaul2011} to measure the information freshness in timely status update systems. In contrast to delay, which measures only the time required to deliver a packet, AoI measures the elapsed time since the latest received update packet was generated. Since AoI is fundamentally different from conventional metrics, it has received significant attention in recent years. Early works on AoI focused on the upper layers of the communication protocol stack \cite{Najm2017,Inoue2017,Kam2018,Arafa2017,Farazi2018,Bedewy2019
}. For example, \cite{Najm2017,Inoue2017,Kam2018} considered different queuing models and analyzed the corresponding average AoI. Scheduling policies that improve information freshness were studied under various network models \cite{Arafa2017,Farazi2018,Bedewy2019
}. Recently, the study of AoI has moved down to the medium access control (MAC) layer and the physical (PHY) layer. For example, AoI with different MAC protocols, including both scheduled access and random access strategies, 
were investigated in \cite{Kuo2020,Yang2020,TTChan2021}. At the PHY layer, \cite{Xie2019,Arafa2019,Li2020} investigated the impacts of channel coding on the average AoI.

In practical wireless systems, packet corruption is inevitable due to wireless impairments. Conventional communication systems are designed for reliable communications, in which Automatic Repeat-reQuest (ARQ) is a practical way to ensure reliability.  The source retransmits the corrupted packet until the destination finally receives the packet. However, when the system metric becomes information freshness, whether ARQ should be used requires re-investigation, because a new packet always contains the most recent information. For example, \cite{Xie2020} studied the average AoI of a point-to-point system and showed that ARQ does not help to reduce the average AoI. Instead, when a source has a chance to send, sampling and sending a new packet (i.e., no ARQ) achieves a higher level of information freshness.

\begin{figure}
\centerline{\includegraphics[trim=0 20 0 0, width=0.3\textwidth]{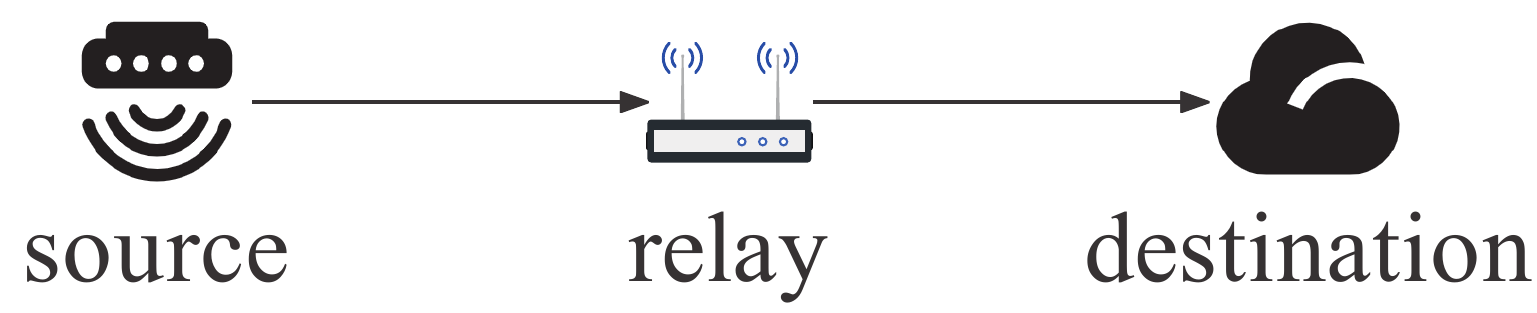}}
\caption{A two-hop status update system with a source node, a relay, and a destination node.}
\label{fig1}
\vspace{-0.1in}
\end{figure}

The study of \cite{Xie2020} was limited to a single hop. In many practical scenarios, however, the destination may be located out of the communication range of the source, e.g., many low-cost sensors may have low transmit power. Thus, a relay is dedicated to helping forward the update packet of the source to the destination, as shown in Fig.~\ref{fig1}. This paper then poses a question: \emph{should ARQ be used in such a two-hop network?}

The answer to the above question is not so obvious. In single-hop networks, the average AoI of the non-ARQ scheme is lower because a new packet always contains the latest status update. If ARQ is not used in two-hop networks, this means that a new packet is sent when the relay fails to receive the packet from the source (i.e., in the first hop), or when the destination fails to receive the packet forwarded by the relay (i.e., in the second hop). By doing so, there is no doubt that the latest status update will be available whenever the destination successfully receives the update packet. However, the time to receive an update could be long, because even if the relay  successfully receives a packet from the source, the packet is immediately discarded if the destination fails to receive it. Sending a new packet at the source can help the destination receive the latest update packet, but the transmission process has to restart from the first hop. Waiting too long for a successful update can result in a high average AoI (i.e., a low information freshness). 

Hence, when decoding fails in a two-hop network, whether to send a new packet at the source or to retransmit the old packet at the relay requires quantitative study to achieve a low average AoI. To this end, this paper presents a theoretical analysis of the average AoI in two-hop networks, considering both the ARQ and the non-ARQ schemes. In particular, we model each scheme using a Markov chain to derive the average AoI. Our theoretical and simulation results show that, unlike single-hop networks, ARQ should be used in two-hop networks to achieve lower average AoI. In particular, our analysis indicates that, for the ARQ scheme, the packet decoding success rate of the second hop has a more significant impact on the average AoI than that of the first hop. 

\begin{figure}
\centerline{\includegraphics[trim=0 30 0 0, width=0.23\textwidth]{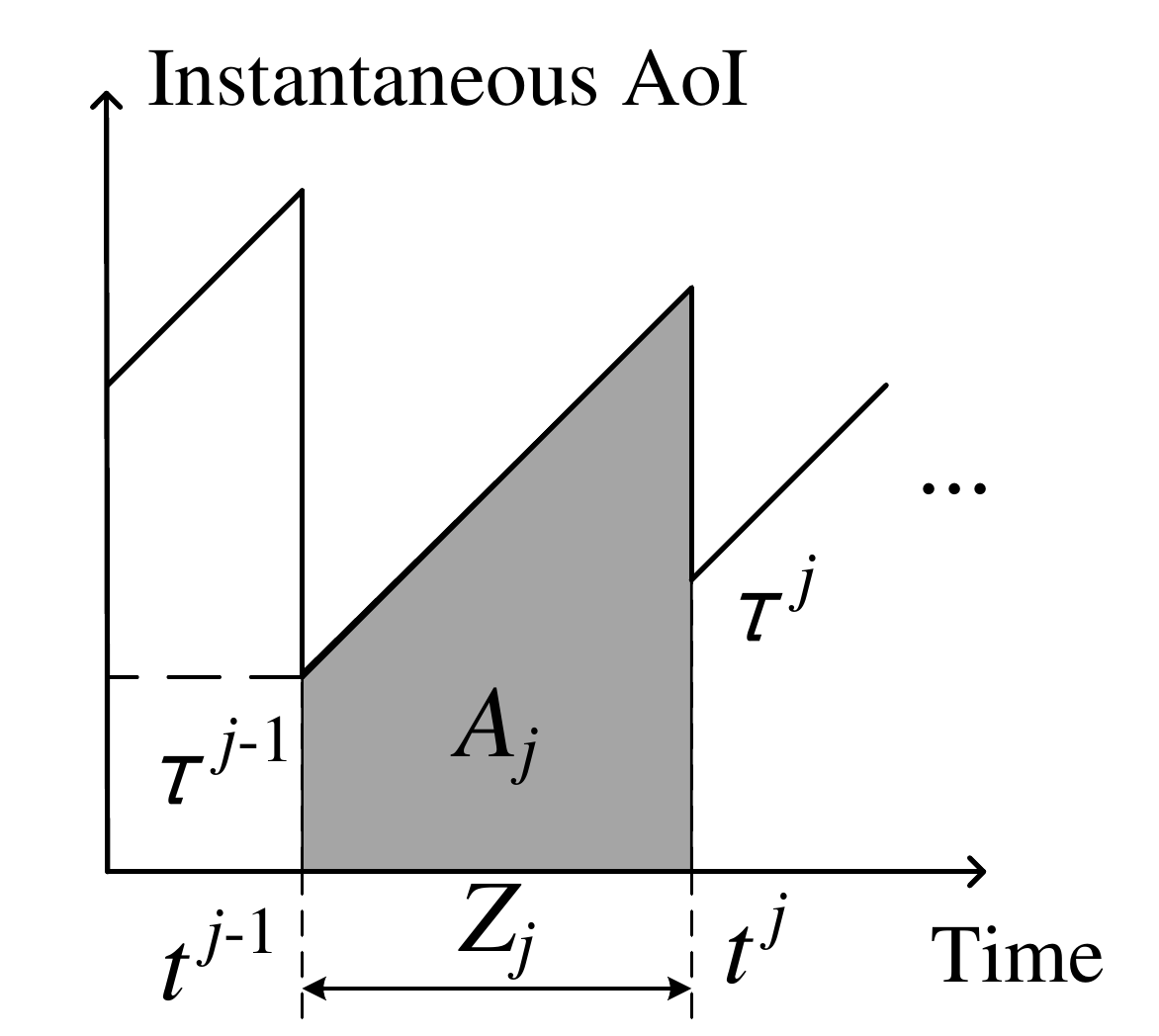}}
\caption{An example of the instantaneous AoI $\Delta (t)$, where the $(j - 1)$-th and the $j$-th successful update occur at times ${t^{j - 1}}$ and ${t^j}$, respectively.}
\label{fig2}
\vspace{-0.1in}
\end{figure}

\section{Preliminaries}
\label{section2}

\subsection{ Age of Information (AoI) Metrics}\label{sec:22}
\label{2.1}

We study a two-hop status update system with a source node, a relay, and a destination node, as shown in  Fig.~\ref{fig1}. The source senses the status of physical characteristics (such as temperature, humidity, etc.) and wants to send status update packets to the destination that is not within the communication range of the source. The decode-and-forward relay helps to forward the update packets of the source to the destination. 

In status update systems, the destination wants to receive update packets from the source as fresh as possible. AoI is used in this paper to quantify the freshness of update packets. Specifically, at any time $t$, the instantaneous AoI of the source measured at the destination is defined by $\Delta (t) = t - G(t)$, where $G(t)$ is the generation time of the most recently received update packet from the source. 

This paper considers a generate-at-will model~\cite{Yates2021}, where the source will take measurements and generate an update packet at any time  it has the opportunity to transmit. Fig.~\ref{fig2} plots an example of the instantaneous AoI $\Delta (t)$, where the $(j - 1)$-th and the $j$-th successful updates occur at times ${t^{j - 1}}$ and ${t^j}$, respectively. As shown in Fig.~\ref{fig2}, the instantaneous AoI $\Delta (t)$ generally increases linearly with time $t$, and drops only when an update packet is successfully decoded by the destination. Denote by $\tau $ the instantaneous AoI at the moment when an update packet is successfully decoded by the destination. Fig.~\ref{fig2} shows that $\Delta (t)$ drops to ${\tau ^{j - 1}}$ and ${\tau ^j}$ at times ${t^{j - 1}}$ and ${t^j}$, respectively.

With the instantaneous AoI $\Delta (t)$, we can compute the average AoI. Specifically, the average AoI $\bar \Delta $ of the source is defined as the time average of the instantaneous AoI
\begin{align}
\bar \Delta  = \mathop {\lim }\limits_{T \to \infty } \frac{1}{T}\int_0^T {\Delta (t)} dt.
\label{f2}
\end{align}

To compute the average AoI $\bar \Delta $, we use $Z$ to represent the time between two consecutive status updates, e.g., Fig.~\ref{fig2} uses ${Z_j}$ to denote the time required for the $j$-th successful update since the $(j - 1)$-th successful update. Let us consider the area $A$ under the line between two consecutive successful updates as shown in Fig.~\ref{fig2}. The area $A_j$ between the $(j-1)$-th and the $j$-th successful updates is calculated by 
\begin{align}
A_j = {\tau ^{j - 1}}{Z_j} + \frac{1}{2}{\left( {{Z_j}} \right)^2}.
\label{f3}
\end{align}
According to the renewal theory, the average AoI $\bar \Delta $ is computed by
\begin{align}
\bar \Delta & = \mathop {\lim }\limits_{J \to \infty } \frac{{\sum\nolimits_{j = 1}^J {{A_j}} }}{{\sum\nolimits_{j = 1}^J {{Z_j}} }} 
= \frac{{E\left[ {\tau Z + \frac{1}{2}{{\left( Z \right)}^2}} \right]}}{{E\left[ Z \right]}} 
= \frac{E\left[ \tau Z  \right]}{E\left[ Z  \right]} + \frac{{E\left[ {{Z^2}} \right]}}{{2E\left[ Z \right]}} .
\label{f4}
\end{align}

As we are considering a status update system with two hops, an important issue is how to deal with the packet loss to ensure a low average AoI of the system. In conventional wireless systems, ARQ is used to ensure reliable transmission at the link layer. However, in terms of AoI, prior works on single-hop networks (i.e., without relays) found that the non-ARQ scheme has a lower average AoI compared with the classical ARQ scheme. The following subsection reviews this finding in single-hop networks, and we believe that it is not directly applicable to the two-hop networks considered in this paper. 

\subsection{Review: The Average AoI of Single-hop Networks}\label{sec:23}
\label{2.2}
We consider a single-hop network in which the source can transmit packets directly to the destination (i.e., without the help of the relay). We assume a time-slotted system in which the transmission time of an update packet occupies a time slot of duration 1 for the sake of simplicity.  

Let us first consider the non-ARQ case, where in each time slot, the source sends a new update packet to the destination, regardless of the decoding result of the destination in the previous time slots. We assume that the destination receives and decodes an update packet successfully with probability $q$. Since a new update packet is sent in each time slot, once the packet is successfully received, the instantaneous AoI of the source drops to $\tau =1$ (i.e., $E[\tau Z ] = E[Z ]$). Furthermore, the time between two consecutive updates, $Z$, is a geometric random variable with parameter $q$, i.e., $E[Z] = 1/q,$ $E[{Z^2}] = (2 - q)/{q^2}$. Hence, substituting the terms into (\ref{f4}), the average AoI of a one-hop network without ARQ, $\bar \Delta _1^{{\rm{N - ARQ}}}$, is \cite{Xie2020}
\begin{align}
\bar \Delta _1^{{\rm{N - ARQ}}} = 1+ \frac{(2 - q)/{q^2}}{2(1/q)}  =\frac{1}{2} + \frac{1}{q}.
\label{f5}
\end{align}

Now let's look at the classic ARQ case.  When ARQ is used, a new update packet is sent only if the destination successfully receives the previous old packet. Notice that the time between two consecutive status updates, $Z$, is still a geometric random variable with the parameter $q$; however, when the update packet is received, the instantaneous AoI of the source now drops to $\tau  = Z'$, where $Z'$ is the corresponding $Z$ in the last update (i.e., $E[\tau Z ] = (E[Z])^2$). According to~\cite{Xie2020}, the average AoI of a one-hop network with ARQ, $\bar \Delta _1^{{\rm{ARQ}}}$, is computed by
\begin{align}
\bar \Delta _1^{{\rm{ARQ}}}& = \frac{1}{q} + \frac{(2 - q)/{q^2}}{2(1/q)}   = \left( {\frac{1}{2} + \frac{1}{q}} \right) + \left( {\frac{1}{q} - 1} \right) \cr
& = \bar \Delta _1^{{\rm{N - ARQ}}} + \left( {\frac{1}{q} - 1} \right).
\label{f6}
\end{align}

Comparing (\ref{f5}) and (\ref{f6}), since $0 < q \le 1$, $\bar \Delta _1^{{\rm{ARQ}}}$ is never smaller than $\bar \Delta _1^{{\rm{N - ARQ}}}$. That is, ARQ does not help to improve the average AoI of single-hop networks. This is because a newer update packet always has the most up-to-date information (i.e., a smaller $\tau$). 

However, we believe that the above conclusion does not hold in two-hop networks. We see from (\ref{f4}) that the average AoI depends on $\tau $ and $Z$. In the two-hop case without ARQ, the minimum instantaneous AoI is $\tau =2$ time slots. In other words, in the absence of ARQ, when an update packet is not successfully received in either the first hop (at the relay) or the second hop (at the destination), the relay will drop the old packet, and the source will immediately send a new update packet. By doing so, as long as the destination receives an update packet, the instantaneous AoI will drop to $\tau =2$.

Although the smallest $\tau $ can be achieved when not using ARQ, blindly transmitting a new update packet from the first hop increases the duration between two consecutive updates, $Z$. For example, the relay may take a long time to receive the latest update packet, but this packet is immediately discarded if the transmission from the relay to the destination fails. By contrast, if ARQ is used in the second hop, the destination may successfully receive the update after only one retransmission from the relay  (i.e., only one more time slot is needed), thus having a smaller $Z$. At the same time, in the ARQ scheme, $\tau $ is larger than two time slots due to packet retransmission. Therefore, a quantitative study is required to thoroughly understand the joint impact of $\tau $ and $Z$ on the average AoI.
\section{The Average AoI of Two-hop Networks}
\label{section3}

\subsection{The Average AoI of the non-ARQ Scheme}
\label{3.1}

We first compute the average AoI of a two-hop network without ARQ. Later, we investigate the average AoI when ARQ is employed in Section \ref{3.2}. Suppose that the relay receives an update packet from the source in the first hop with a successful decoding probability of ${p_1}$, and the destination receives an update packet from the relay in the second hop with a successful decoding probability of ${p_2}$.

\begin{figure}
\centerline{\includegraphics[trim=0 30 0 0, width=0.45\textwidth]{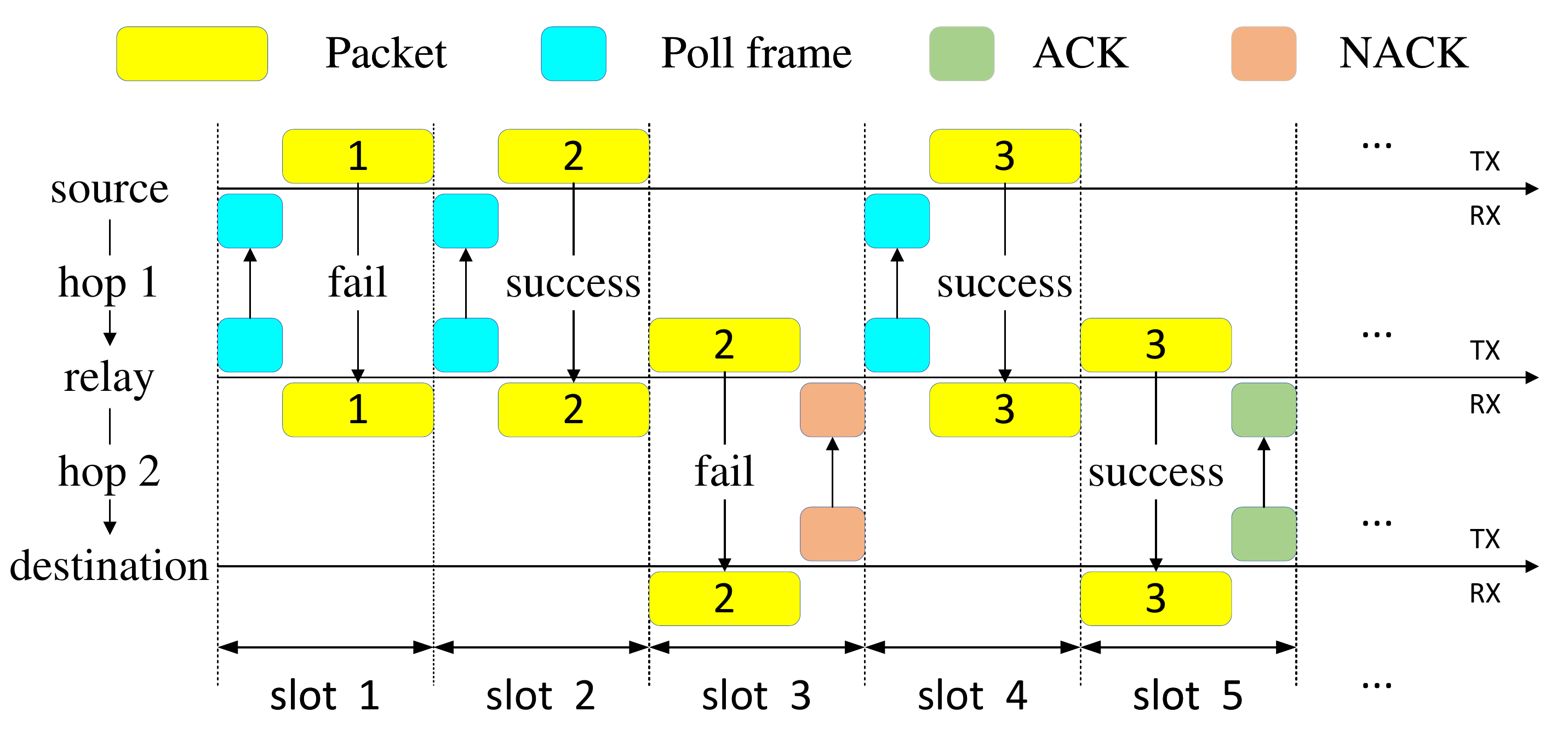}}
\caption{The MAC protocol in a two-hop network without ARQ.}
\label{fig3}
\vspace{-0.15in}
\end{figure}

Fig.~\ref{fig3} depicts an example of the MAC protocol without ARQ in a two-hop network. The relay first sends a polling frame to the source. After receiving the polling frame, the source samples and sends an update packet (i.e., packet $1$) to the relay. Suppose that the relay cannot decode the update packet. The relay informs the source to send a new update packet by sending a polling frame again, and now the update packet (i.e., packet $2$) is successfully decoded by the relay. After that, the relay forwards the update packet to the destination, but the destination cannot decode the packet and sends a negative acknowledgment (NACK) frame to the relay. Since ARQ is not used, the relay simply drops the old packet and informs the source to send a new packet (i.e., packet $3$). Fig.~\ref{fig3} assumes that both the relay and the destination receive packet $3$, and then the destination sends an acknowledgment (ACK) frame to the relay.

\begin{figure}
\centerline{\includegraphics[trim=0 30 0 0, width=0.35\textwidth]{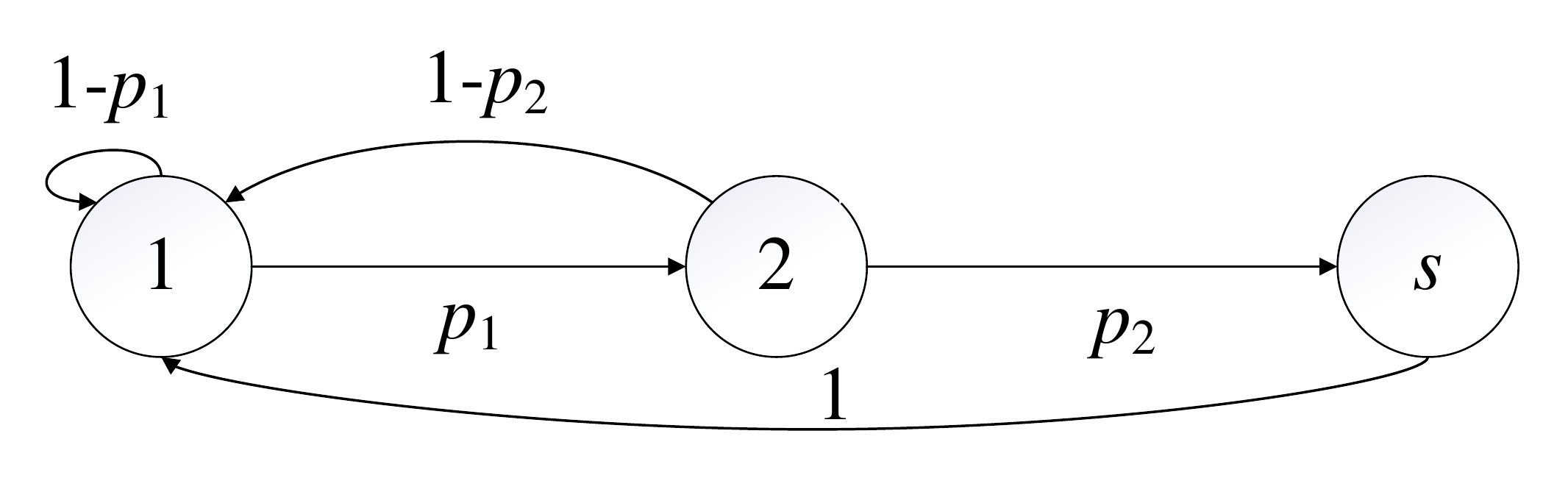}}
\caption{The Markov Chain of the non-ARQ scheme in a two-hop network, where state $1$ $(2)$ means that the current transmission is in the first (second) hop, and state $s$ indicates a successful update at the destination.}
\label{fig4}
\vspace{-0.2in}
\end{figure}

We further assume that the duration of control frames (i.e., the polling frame and the ACK/NACK frame) is negligible compared with that of an update packet. For example, in IEEE 802.11 \cite{IEEE2016}, the duration of a packet with a 512-byte payload is 778 $\mu$s, while the duration of an ACK is only 60 $\mu$s. Therefore, we can still use the time slot as the unit when calculating the average AoI as in Section \ref{2.2}.

We model the MAC protocol using a Markov chain, as shown in Fig.~\ref{fig4}. Specifically, state $1$ $(2)$ means that the current transmission is in the first (second) hop, and state $s$ indicates a successful update at the destination. We use $Q$ to denote the state space, i.e., $Q = \{ 1,2,s\} $. At the beginning, the system starts with state $1$, i.e., the first hop. A new update packet is generated and sent to the relay. The relay can decode the update packet from the source with a  probability of ${p_1}$. If the packet is successfully decoded, the Markov chain will transit to state $2$, meaning that the next time slot is used for the second hop transmission. Otherwise, if the relay cannot decode the update packet from the source (probability: $1-{p_1}$), the system remains in state $1$. That is, the next time slot is still used for the communication of the first hop, and the source drops the old packet and sends a new packet to the relay.

When the current state is state $2$, the relay forwards the update packet to the destination. If the destination successfully receives the packet  (probability: ${p_2}$), the Markov chain transits to state $s$, meaning that the destination successfully receives the update, i.e., the instantaneous AoI drops to $2$ time slots and $E\left[ \tau Z  \right] = 2E\left[ Z  \right]$. Otherwise, with probability $1-{p_2}$, the destination cannot decode the update packet, and the state goes back to state $1$. Finally, when the current state is $s$, the next state must be state $1$ since a new update packet will be sent, starting from the first hop. 
Let ${\Pi ^{N - ARQ}}$ denote the state transition matrix, which can be written as 
\begin{align}
{\Pi ^{N - ARQ}} =\begin{pmatrix}
    {{\pi _{11}}} & {{\pi _{12}}} & {{\pi _{1s}}}  \cr 
    {{\pi _{21}}} & {{\pi _{22}}} & {{\pi _{2s}}}  \cr 
    {{\pi _{s1}}} & {{\pi _{s2}}} & {{\pi _{ss}}}  \cr 
\end{pmatrix}=\begin{pmatrix}
    1 - {p_1} & {p_1} & 0  \cr 
    1 - {p_2} & 0 & {p_2}  \cr 
     1 & 0 & 0  \cr  
\end{pmatrix}
\label{f7}
\end{align}
where ${\pi _{xy}}$ is the probability of transiting from state $J = x$ to state $J = y$, for $x,y \in Q$.

To compute the average AoI of the non-ARQ scheme by (\ref{f4}), we first compute $E\left[ Z \right]$ and $E\left[ {{Z^2}} \right]$. We use ${m_{is}}$ to represent the expected time required to transverse from state ${J_0} = i$ to state ${J_Z} = s$ for the first time through a series of states ${J_1}, {J_2}, \dots, {J_{Z-1}}$. Based on the property of the Markov chain, we have
\begin{align}
{m_{is}}& =  E\left[ {{T_s}\left| {{J_0} = i} \right.} \right] \cr
& =\left\{
\begin{aligned}
& 0 &,\; & i = s, \\
& \sum\limits_{j \in Q} {E\left[ {{1+T_s}\left| {{J_1} = j} \right.} \right]P\left( {{J_1} = j\left| {{J_0} = i} \right.} \right)} &, \;& i \ne s,
\end{aligned}
\right. \cr
 & =\left\{
\begin{aligned}
& 0 &,\; & i = s, \\
& 1 + \sum\limits_{j \ne s} {{\pi _{ij}}{m_{js}}} &, \;& i \ne s,
\end{aligned}
\right.
\label{f8}
\end{align}
where ${T_s}$ is a random variable that represents the time to reach state $J = s$ for the first time. According to (\ref{f8}), we have
\begin{align}\left\{
\begin{aligned}
& {m_{1s}} = 1 + {p_1}{m_{2s}} + (1 - {p_1}){m_{1s}}, \cr
& {m_{2s}} = 1  + (1 - {p_2}){m_{1s}}, \cr
& {m_{ss}} = 0.
\end{aligned}
\right.
\label{f9}
\end{align}
Based on the Markov chain shown in Fig.~\ref{fig4}, $E\left[ Z \right]$ equals ${m_{1s}}$. By simplifying  (\ref{f9}), $E\left[ Z \right]$ can be computed by
\begin{align}
E\left[ Z \right] = {m_{1s}} = \frac{{1 + {p_1}}}{{{p_1}{p_2}}} = \frac{1}{{{p_1}{p_2}}} + \frac{1}{{{p_2}}}.
\label{f10}
\end{align}
Similarly, we use ${n_{is}}$ to denote the expectation of the second moment of the time required to transverse from state from state ${J_0} = i$  to state ${J_Z} = s$ for the first time through a series of states ${J_1}, {J_2}, \dots, {J_{Z-1}}$. By definition, ${n_{is}}$ is computed by
\begin{align}
&{n_{is}} =  E\left[ {{{\left( {{T_s}} \right)}^2}\left| {{J_0} = i} \right.} \right]   \cr
&=\left\{\begin{aligned}
& 0 & , \;& i = s, \\
& \sum\limits_{j \in Q} {E\left[ {{{\left( {1 + {T_s}} \right)}^2}\left| {{J_1} = j} \right.} \right]P\left( {{J_1} = j\left| {{J_0} = i} \right.} \right)} &, \;& i \ne s.
\end{aligned}
\right.
\nonumber 
\end{align}
\begin{align}
 &=\left\{\begin{aligned}
& 0 & , \;& i = s, \\
& \sum\limits_{j \in Q} {\left( \begin{array}{l}
E\left[ {{{\left( {{T_s}} \right)}^2}\left| {{J_1} = j} \right.} \right]P\left( {{J_1} = j\left| {{J_0} = i} \right.} \right)\cr
 + 2E\left[ {{T_s}\left| {{J_1} = j} \right.} \right]P\left( {{J_1} = j\left| {{J_0} = i} \right.} \right)\cr
 + E\left[ {1\left| {{J_1} = j} \right.} \right]P\left( {{J_1} = j\left| {{J_0} = i} \right.} \right)
\end{array} \right)} &, \;& i \ne s.
\end{aligned}
\right.
\cr
&=\left\{\begin{aligned}
& 0 & , \;& i = s, \\
& 1 + \sum\limits_{j \ne s} {{\pi _{ij}}\left( {{n_{js}} + 2{m_{js}}} \right)} &, \;& i \ne s.
\end{aligned}
\right.
\label{f11}
\end{align}
Therefore, we have
\begin{align}\left\{
\begin{aligned}
& {n_{1s}} = 1+{p_1}\left( {{n_{2s}} + 2{m_{2s}}} \right) + (1 - {p_1})\left( {{n_{1s}} + 2{m_{1s}}} \right), \cr
& {n_{2s}} = 1 + (1 - {p_2})\left( {{n_{1s}} + 2{m_{1s}}} \right), \cr
& {n_{ss}} = 0.
\end{aligned}
\right.
\label{f12}
\end{align}
It is easy to see that $E\left[ {{Z^2}} \right]$ equals ${n_{1s}}$, which is given by
\begin{align}
E\left[ {{Z^2}} \right] = {n_{1s}} = \frac{{2{{\left( {1 + {p_1}} \right)}^2}}}{{{{\left( {{p_1}{p_2}} \right)}^2}}} - \frac{1}{{{p_2}}} - \frac{3}{{{p_1}{p_2}}}.
\label{f13}
\end{align}
With $E\left[ Z \right]$ and $E\left[ {{Z^2}} \right]$, the average AoI of two-hop networks without ARQ, $\bar \Delta _{\rm{2}}^{{\rm{N - ARQ}}}$,  is computed by
\begin{align}
\bar \Delta _{\rm{2}}^{{\rm{N - ARQ}}} = \frac{E\left[ \tau Z  \right]}{E\left[ Z  \right]} + \frac{{E\left[ {{Z^2}} \right]}}{{2E\left[ Z \right]}} = \frac{3}{2} + \frac{{1 + {p_1}}}{{{p_1}{p_2}}} - \frac{1}{{1 + {p_1}}}.
\label{f14}
\end{align}

\subsection{The Average AoI of the ARQ Scheme}
\label{3.2}
\begin{figure}
\centerline{\includegraphics[trim=0 40 0 0, width=0.5\textwidth]{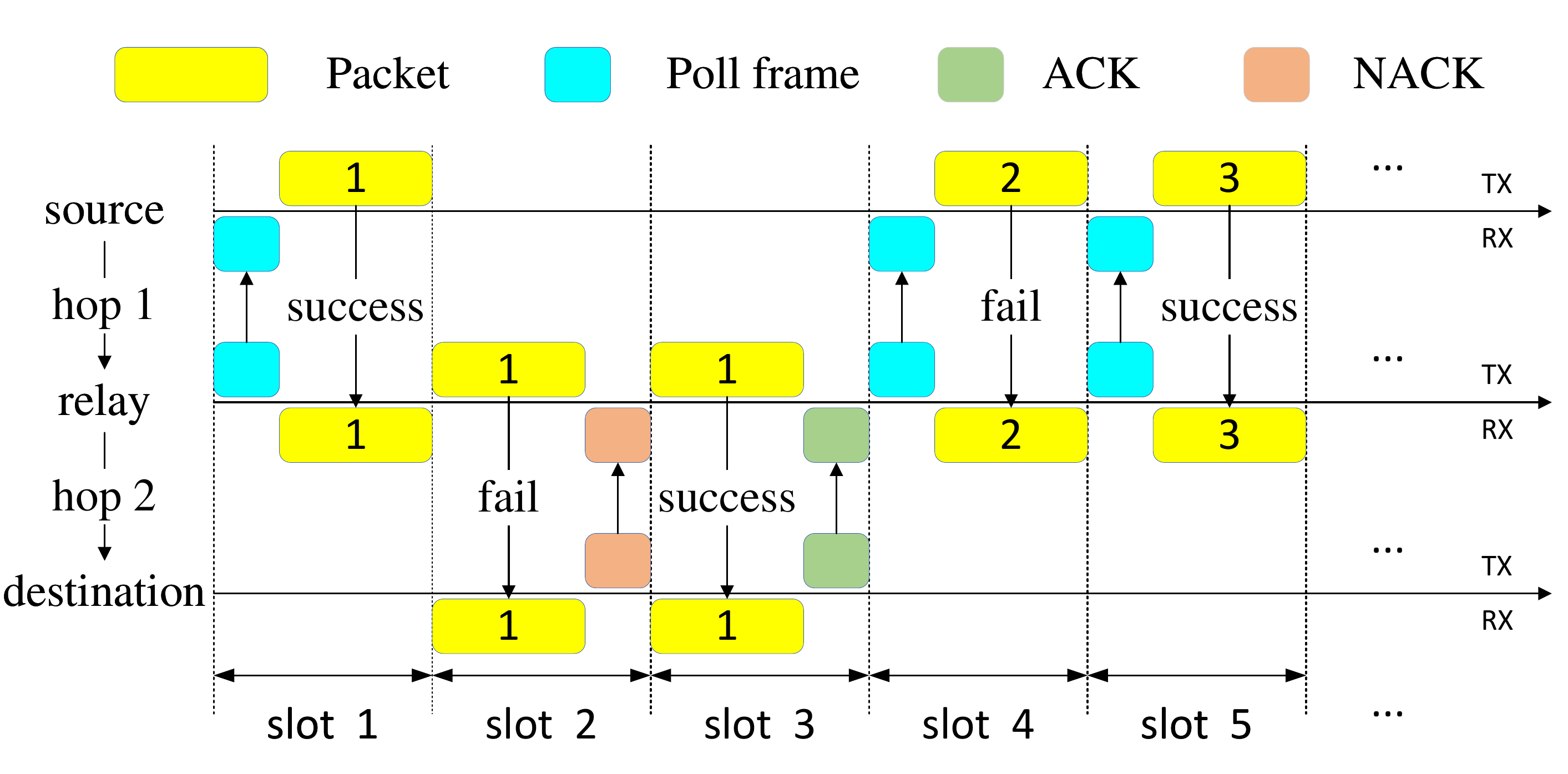}}
\caption{The MAC protocol in a two-hop network with ARQ.}
\label{fig5}
\vspace{-0.2in}
\end{figure}

This subsection analyzes the average AoI of two-hop networks with ARQ. Fig.~\ref{fig5} depicts the MAC protocol in a two-hop network with ARQ. As shown in Fig.~\ref{fig5}, in time slot $2$, when the relay fails to forward packet $1$ to the destination, the destination sends a NACK frame to the relay. The relay retransmits the old packet to the destination in slot $3$. Furthermore, we remark here that ARQ is used only in the second hop. For example, as shown in Fig.~\ref{fig5}, when the relay fails to receive packet $2$ in the first hop in slot $4$, the source sends a new packet (i.e., packet $3$) instead of retransmitting the old packet $2$. In this case, it is easy to understand that sending a new packet from the source (i.e., without ARQ) always leads to a lower instantaneous AoI compared with sending the old packet from the source.

\begin{figure}
\centerline{\includegraphics[trim=0 30 0 0, width=0.38\textwidth]{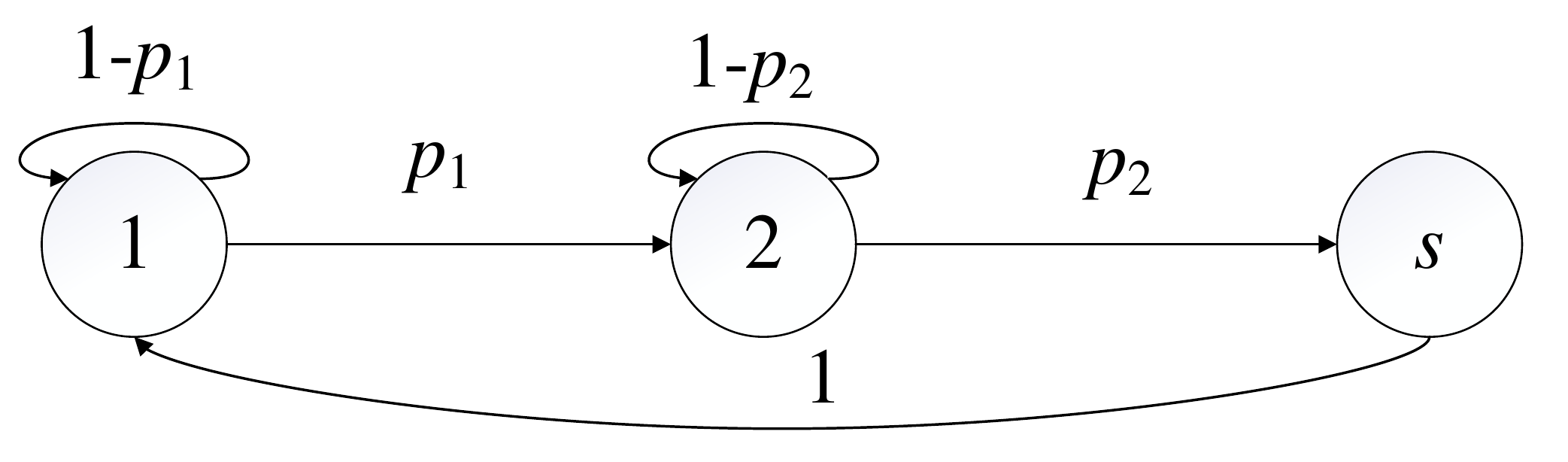}}
\caption{The Markov Chain of the ARQ scheme in a two-hop network, where state $1$ $(2)$ means that the current transmission is in the first (second) hop, and state $s$ indicates a successful update at the destination.}
\label{fig6}
\vspace{-0.2in}
\end{figure}

We also use a Markov chain to model the ARQ scheme in a two-hop network, as shown in Fig.~\ref{fig6}. The only difference compared with the non-ARQ scheme shown in Fig.~\ref{fig4}  is that when the relay fails to forward the update packet to the destination (probability: $1-{p_2}$), the system remains in state $2$ in the ARQ scheme. In contrast, in the non-ARQ scheme, the system transits to state $1$, as shown in Fig.~\ref{fig4}. 

As in the non-ARQ scheme, we need to compute $E\left[ Z \right]$ and $E\left[ {{Z^2}} \right]$ based on the Markov chain. Specifically, We have
\begin{align}\left\{
\begin{aligned}
& {m_{1s}} = 1 + {p_1}{m_{2s}} + (1 - {p_1}){m_{1s}}, \\
& {m_{2s}} = 1 +  (1 - {p_2}){m_{2s}}. 
\end{aligned}
\right.
\label{f16}
\end{align}
Then, $E\left[ {{Z}} \right]$ can be found by
\begin{align}
E\left[ Z \right] = {m_{1s}} = \frac{1}{{{p_1}}} + \frac{1}{{{p_2}}}.
\label{f17}
\end{align}
Similarly, ${n_{is}}$ and $E\left[ {{Z^2}} \right]$ can be computed by
\begin{align}&\left\{
\begin{aligned}
& {n_{1s}} = 1+{p_1}\left( {{n_{2s}} + 2{m_{2s}}} \right) + (1 - {p_1})\left( {{n_{1s}} + 2{m_{1s}}} \right), \\
& {n_{2s}} = 1 + (1 - {p_2})\left( {{n_{2s}} + 2{m_{2s}}} \right),
\end{aligned}
\right.
\label{f18}
\\
&E\left[ {{Z^2}} \right] = {n_{1s}} = \frac{2}{{{{\left( {{p_1}} \right)}^2}}} + \frac{2}{{{{\left( {{p_2}} \right)}^2}}} + \frac{2}{{{p_1}{p_2}}} - \left( {\frac{1}{{{p_1}}} + \frac{1}{{{p_2}}}} \right).
\label{f19}
\end{align}
Let $Y$ denote the time it takes for the relay to receive an update packet from the source, and $X$ denote the time it takes for the destination to receive an update packet from the relay. Then $X$ is a geometric random variable with parameter ${p_2}$. Since ARQ is not used in the first hop and the time taken from the source to the relay is always one time slot, the instantaneous AoI $\tau $ upon a successful update is $\tau  = 1 + X$. Thus, we have
\begin{align}
E\left[ \tau Z  \right] &= E\left[ (1+X')(Y+X)  \right] \notag \\
&= E\left[ Y \right] +E\left[ X \right]+ (E\left[ X \right])^2 +E\left[ Y \right]E\left[ X \right] \notag \\
&=(1+E\left[ X \right])E\left[ Z \right]
\label{f20}
\end{align}
where $X'$ is the corresponding $X$ in the last update. The average AoI of two-hop networks with ARQ, $\bar \Delta _{\rm{2}}^{{\rm{ARQ}}}$, is
\begin{align}
\bar \Delta _{\rm{2}}^{{\rm{ARQ}}} &= \frac{E\left[ \tau Z  \right]}{E\left[ Z  \right]}  + \frac{{E\left[ {{Z^2}} \right]}}{{2E\left[ Z \right]}} =\frac{1}{2} + \frac{1}{{{p_1}}} + \frac{2}{{{p_2}}} - \frac{1}{{{p_1} + {p_2}}}.
\label{f21}
\end{align}

Comparing the average AoI performance of the ARQ scheme (\ref{f21}) and the non-ARQ scheme (\ref{f14}), we have
\begin{align}
&\bar \Delta _{\rm{2}}^{{\rm{ARQ}}} - \bar \Delta _{\rm{2}}^{{\rm{N - ARQ}}} \nonumber \\ = &\frac{{\left( {1 - {p_2}} \right)\left\{ {\left[ {{{\left( {{p_1}} \right)}^2} - 1} \right]\left( {{p_1} + {p_2}} \right) - {p_1}{p_2}} \right\}}}{{\left( {{p_1} + {p_2}} \right)\left( {{p_1}{p_2}} \right)\left( {1 + {p_1}} \right)}}\leq0,
\label{f22}
\end{align}
because ${p_1},{p_2}\in [0,1]$. As a result, we can conclude that $\bar \Delta _2^{{\rm{ARQ}}} \le \bar \Delta _2^{{\rm{N - ARQ}}}$. Recall that in a one-hop network, the average AoI of the non-ARQ scheme is lower than or equal to that of the ARQ scheme (i.e., $\bar \Delta _1^{{\rm{N - ARQ}}} \le \bar \Delta _1^{{\rm{ARQ}}}$; see (\ref{f6})). Interestingly, here we see that $\bar \Delta _2^{{\rm{ARQ}}} \le \bar \Delta _2^{{\rm{N - ARQ}}}$ in the two-hop network. In other words, from the perspective of enhancing the information freshness, unlike one-hop networks, ARQ should be used in two-hop networks. In the next section, we further compare the simulation results of $\bar \Delta _2^{{\rm{N - ARQ}}}$ and $\bar \Delta _2^{{\rm{ARQ}}}$ under different ${p_1}$ and ${p_2}$.

\section{Performance Evaluation}
\label{section4}
We now compare the average AoI of the two-hop network with and without ARQ under different ${p_1}$ and ${p_2}$. Specifically, we validate our theoretical analysis via simulations on MATLAB. As we will see, unlike the single-hop network, the ARQ scheme leads to a lower average AoI than the non-ARQ scheme does in the two-hop network.

We consider both the theoretical results and the simulation results. For the theoretical results, we substitute ${p_1}$ and ${p_2}$ into (\ref{f14}) and (\ref{f21}). For the simulation results, we first simulate the two protocols based on ${p_1}$ and ${p_2}$ over a series of time slots and collect the instantaneous AoI in each time slot. After that, we compute the average AoI based on the instantaneous AoI.

\subsubsection{${p_1}={p_2}$ Case} 
\begin{figure}
\centerline{\includegraphics[trim=0 30 0 0, width=0.37\textwidth]{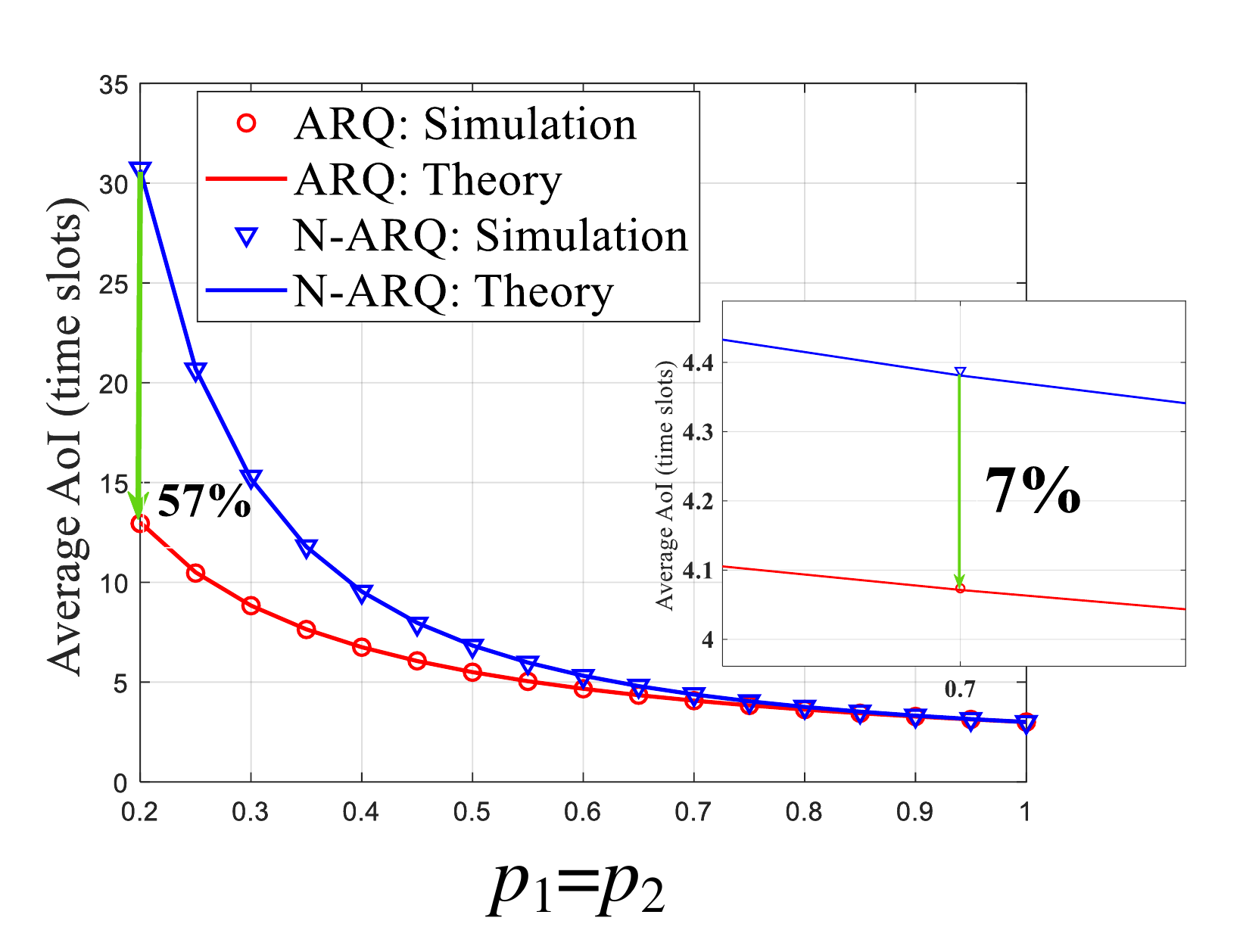}}
\caption{Performance comparison between ARQ and non-ARQ schemes in a two-hop network: the average AoI versus the probability of successful transmission in each hop with ${p_1}$=${p_2}$.}
\label{fig7}
\vspace{-0.15in}
\end{figure}

We examine the relationship between the average AoI and the successful transmission probabilities, when ${p_1}$ is equal to ${p_2}$, as shown in Fig.~\ref{fig7}. We see from Fig.~\ref{fig7} that the simulation results corroborate the theoretical results. Specifically, the average AoI with ARQ is smaller than that without ARQ, as also indicated by (\ref{f22}) previously. This is because, in the non-ARQ scheme, when the second hop transmission fails, the transmission has to be restarted from the first hop. Blindly transmitting a new update packet from the source results in a longer time between two consecutive status updates, i.e., $\frac{1}{{{p_1}}} + \frac{1}{{{p_2}}}$ for the ARQ scheme and $\frac{1}{{{p_1}{p_2}}} + \frac{1}{{{p_2}}}$ for the non-ARQ scheme. Even though the non-ARQ scheme has a smaller instantaneous AoI upon a successful update (i.e., two time slots), the ARQ scheme still has a lower average AoI due to the shorter ``inter-update'' interval.

Furthermore, the ARQ scheme reduces the average AoI more significantly when ${p_1}$ and ${p_2}$ are small. In particular, when ${p_1}={p_2}=0.2$, the average AoI is reduced by around $57\%$ when using the ARQ scheme compared with the non-ARQ scheme. As ${p_1}$ and ${p_2}$ increase, we find that the performance improvement from ARQ becomes smaller because there are fewer packet corruptions (i.e., ARQ is not often needed, so the average AoI are almost the same for both schemes when ${p_1}$ and ${p_2}$ are larger). For example, when ${p_1}={p_2}=0.7$, the average AoI drops by about $7\%$ for the ARQ scheme compared with the non-ARQ scheme.

\subsubsection{${p_1}\ne{p_2}$ Case} 
\begin{figure}
\centerline{\includegraphics[trim=0 30 0 0, width=0.5\textwidth]{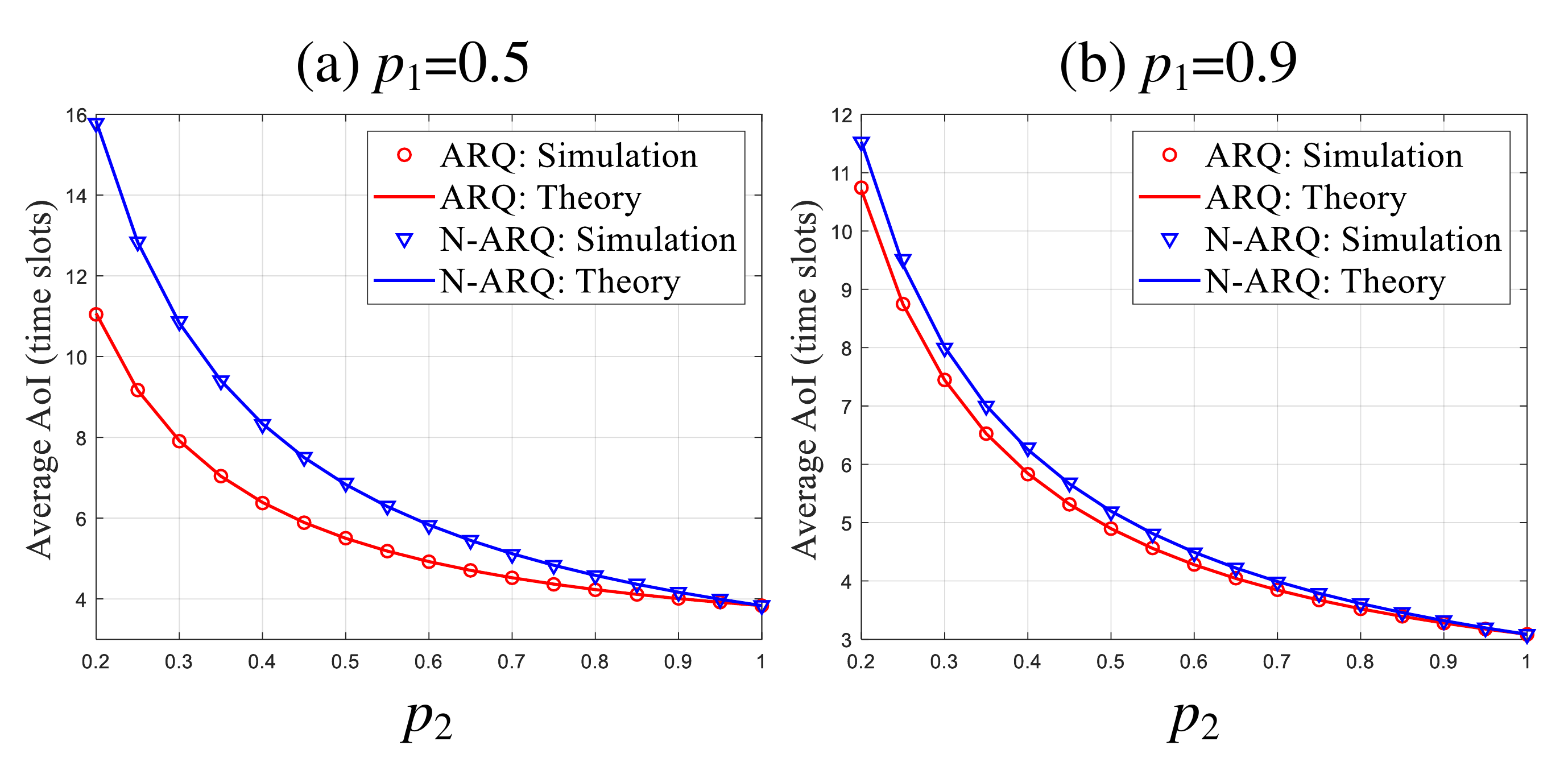}}
\caption{Performance comparison between the ARQ scheme and the non-ARQ scheme: the average AoI versus ${p_2}$, when ${p_1}$ is fixed to (a) $0.5$ and (b) $0.9$.}
\label{fig8}
\vspace{-0.2in}
\end{figure}

\begin{figure}
\centerline{\includegraphics[trim=0 30 0 0, width=0.5\textwidth]{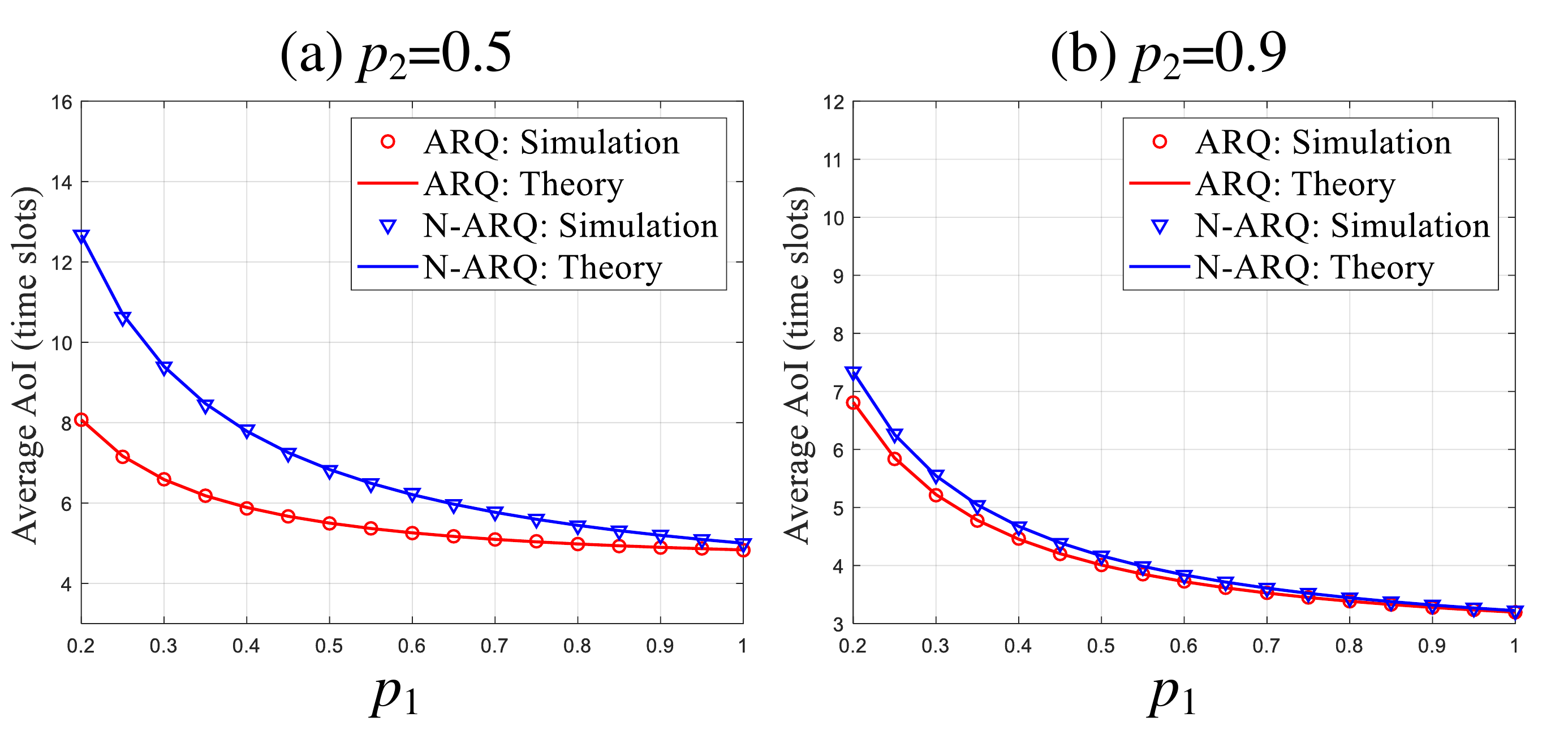}}
\caption{Performance comparison between the ARQ scheme and the non-ARQ scheme: the average AoI versus ${p_1}$, when ${p_2}$ is fixed to (a) $0.5$ and (b) $0.9$.}
\label{fig9}
\vspace{-0.2in}
\end{figure}

We next focus on the case where ${p_1} \ne {p_2}$. Specifically, Fig.~\ref{fig8} plots the average AoI versus ${p_2}$, when ${p_1}$ is fixed to (a) $0.5$ and (b) $0.9$; Fig.~\ref{fig9} plots the average AoI versus ${p_1}$, when ${p_2}$ is fixed to (a) $0.5$ and (b) $0.9$. From Fig.~\ref{fig8} and Fig.~\ref{fig9}, we find that, as in the case of ${p_1}={p_2}$, the average AoI of the two-hop network with ARQ is smaller than that without ARQ, i.e., $\bar \Delta _2^{{\rm{ARQ}}} \le \bar \Delta _2^{{\rm{N - ARQ}}}$. In addition, the ARQ scheme reduces the average AoI more significantly when ${p_1}$ and ${p_2}$ are small.

Moreover, when ARQ is used, ${p_2}$ has a greater impact on the average AoI than ${p_1}$ does. For example, as shown in Fig. 8(a), when ${p_1}=0.5$ and ${p_2}$ varies from $0.2$ to $1$, the average AoI of the ARQ scheme drops by $65.4\%$ from $11.07$ time slots to $3.83$ time slots. However, as shown in Fig. 9(a), when ${p_2}=0.5$ and ${p_1}$ varies from $0.2$ to $1$, the average AoI drops by $40.1\%$ from $8.07$ time slots to $4.83$ time slots. Similar effects can be found when we compare Fig. 8(b) and Fig. 9(b) with a larger $p_1$ or $p_2$. Furthermore, let us consider two special cases where $(p_1,p_2)=(0.9,0.5)$ and $(p_1,p_2)=(0.5,0.9)$. It is easy to figure out that $(p_1,p_2)=(0.5,0.9)$ leads to a lower average AoI. This can be explained by (\ref{f21}), the average AoI of the ARQ scheme. Specifically, (\ref{f21}) can be decomposed into two parts, i.e., $\frac{1}{2} + \frac{{{p_1} + {p_2}}}{{{p_1}{p_2}}} - \frac{1}{{{p_1} + {p_2}}}$ and $\frac{1}{{{p_2}}}$. It is easy to observe that swapping the two hops' successful rates does not affect the first part of the equation, while a higher $p_2$ leads to a lower $\frac{1}{{{p_2}}}$ and hence a lower average AoI. 

\section{Conclusion}
We have compared the average AoI performances between the ARQ and the non-ARQ schemes in two-hop networks. We derive the theoretical average AoI of both schemes using Markov chains. Unlike single-hop networks in which ARQ does not help to improve information freshness, our theoretical and simulation results indicate that ARQ should be used in two-hop networks to lower the average AoI. Moreover, we find that the packet decoding success rate of the second hop has a more significant impact on the average AoI than that of the first hop does. We believe that the insights of ARQ designs in two-hop networks with AoI requirements are generally applicable to multi-hop line networks beyond two hops, whereas the detailed investigation is left to our future work.


\end{document}